\documentclass[%
 aip,
 amsmath,amssymb,
 reprint,%
]{revtex4-1}

\usepackage{graphicx}
\usepackage{dcolumn}
\usepackage{bm}

\usepackage[utf8]{inputenc}
\usepackage[T1]{fontenc}
\usepackage{mathptmx}
\usepackage{etoolbox}

\makeatletter
\def\@email#1#2{%
 \endgroup
 \patchcmd{\titleblock@produce}
  {\frontmatter@RRAPformat}
  {\frontmatter@RRAPformat{\produce@RRAP{*#1\href{mailto:#2}{#2}}}\frontmatter@RRAPformat}
  {}{}
}%
\makeatother
\begin{document}

\preprint{AIP/123-QED}

%
%
%
%

\title{Comparison of DBS measurements of turbulence spectra by vertical-displacement and poloidal-angle scans using the Scotty synthetic diagnostic}

\author{Y.T. Tan}
\email{Tan\_Yunn\_Ting@a-star.edu.sg}
\affiliation{Future Energy Acceleration and Translation Centre, Agency for Science, Technology and Research (A*STAR), Singapore 138632, Singapore}

\author{V.H. Hall-Chen}
\affiliation{Future Energy Acceleration and Translation Centre, Agency for Science, Technology and Research (A*STAR), Singapore 138632, Singapore}
\affiliation{School of Physical and Mathematical Sciences, Nanyang Technological University, Singapore 639798, Singapore}

\author{T.L. Rhodes}
\affiliation{Physics and Astronomy Department, University of California, Los Angeles (UCLA), Los Angeles, California 90095, USA}

\date{\today} 

\begin{abstract}
Doppler backscattering (DBS) measures electron density fluctuations. The measured wavenumber is typically varied by changing the probe beam’s poloidal launch angle. As most DBS systems are unable to steer during a shot, the shot is repeated and the poloidal angle is changed intershot. An alternative method is to keep the DBS launch angle fixed and move the plasma up and down instead, enabling a range of wavenumbers to be measured within a single shot. We call this the bouncing ball method. We use the Scotty synthetic diagnostic (Hall-Chen, 2022) to evaluate the similarities and differences between these two approaches. Both approaches are capable of measuring a similar range of fluctuation wavenumbers as well as radial locations. However, the vertical-displacement scan has a larger range of measured poloidal locations than the poloidal-angle scan. Using the same synthetic turbulence spectrum as input, we show that the two approaches are expected to have different backscattered powers due to different instrumentation functions. When mismatch attenuation is accounted for via a synthetic diagnostic, the vertical-displacement scan can provide comparable radial and wavenumber coverage while reducing reliance on shot-to-shot repeatability. These results establish vertical-displacement scan as a practical route to single-shot DBS wavenumber spectra. 
\end{abstract}

\maketitle

\section{Introduction}
Turbulent transport is the dominant mechanism of heat and particle transport in tokamak plasmas \cite{Garbet:turbulence:2010}. As such, understanding how to suppress turbulence and how this would extrapolate to future fusion devices is an important open problem \cite{White:validation:2019}. The Doppler backscattering (DBS) diagnostic\cite{Rhodes:DBS:2022, Liu:DBS:2023, Shi:DBS:2023, Cabrera:DBS:2023, Damba:DBS:2024, Liang:DBS:2026} can measure turbulent density fluctuations of intermediate scales, $1 \lesssim k_\perp \rho_s \lesssim 10$. Here $k_\perp$ is the perpendicular wavenumber of the turbulent density fluctuations and $\rho_s$ is the ion sound gyroradius. The backscattered signal is proportional to the amplitude of the density fluctuations. Moreover, DBS has good spatial and temporal resolution. Hence, DBS is well-suited to measure density fluctuations and validate gyrokinetic models of turbulence.

Measuring the density fluctuation spectrum is complicated. First, the poloidal launch angle is typically varied, such that the DBS probe beam reaches approximately the same cutoff locations but with a different wavenumber. By the Bragg condition, the fluctuation wavenumber responsible for backscattering is twice that of the probe beam's wavenumber. By varying the poloidal launch angle systematically, typically shot-to-shot, the backscattered power spectrum is measured\cite{Hennequin:DBS:2006, Hillesheim:DBS:2015, Happel:DBS:2017, Estrada:DBS:2019, Pratt:DBS:2024}. Secondly, inverting the backscattered power spectrum to recover the density fluctuation spectrum is non-trivial. One approach is to simulate the density fluctuations and use a synthetic diagnostic to predict the backscattered spectral density, then compare the predicted and measured spectral densities directly\cite{Happel:DBS:2017, Pratt:DBS:2024}. It is worth noting that, apart from DBS, the density fluctuation spectrum can also be measured by phase-contrast imaging\cite{Tanaka:PCI:2008, Kinoshita:PCI:2023}, short-pulse reflectometry\cite{Krutkin:SPR:2024}, and high-k scattering\cite{Barchfeld:highk:2018, Sun:highk:2024, Speirs:highk:2025}.

In this paper, we propose an alternative method to measure the backscattered spectral density with DBS. Instead of scanning the poloidal launch angle, either the DBS antenna or the plasma can be translated up and down instead. In DIII-D, which has good control systems, limited plasmas may be translated up and down within a single shot, enabling the backscattered power spectrum to be measured at multiple locations simultaneously. Such \emph{bouncing ball} plasmas were achieved on DIII-D, and our analysis of the electron temperature and density, ion temperature, and magnetic equilibria indicates that the plasma remains sufficiently similar as it is vertically translated. Such in-shot vertical-displacement scan is more efficient than the typical method of repeating shots, changing the poloidal launch angle after each shot. While in-shot poloidal steering is possible for some DBS systems\cite{Chowdhury:DBS:2023}, it is not common. Conversely, at tokamaks like J-TEXT where the port window has a large vertical extent, moving the DBS antenna up and down between shots could be easier to implement than poloidal steering.

We then used the \textit{Scotty} beam-tracing code\cite{Hall-Chen:DBS_beammodel:2022} as the synthetic diagnostic to compare vertical-displacement and poloidal-angle scans. Given the plasma equilibrium, a form for the turbulent density fluctuations, and properties of the launch beam, \textit{Scotty} then calculates the cutoff location, dominant measured fluctuation wavenumber, and synthetic backscattered signal. This allows us to directly compare the diagnostic response of the two scan methods.

The rest of the paper is structured as follows. In section \ref{sec:methodology}, we introduce the equilibrium and synthetic diagnostic used in this study. We then compare DBS measurements of density fluctuations via vertical-displacement and poloidal-angle scans in section \ref{sec:results}.

\section{Plasma scenario and the \textit{Scotty} synthetic diagnostic} \label{sec:methodology} 
In this section, we detail the methodology used to compare vertical-displacement and poloidal-angle scans. We first describe the DIII-D equilibrium and DBS system used in the calculations. The nature of the two scans are then detailed and explained. With the input equilibrium and parameters scans defined, we then describe how the \textit{Scotty} synthetic diagnostic is used to calculate cutoff locations, measured fluctuation wavenumbers, and ultimately, the synthetic backscattered power.

We considered a limited L-mode DIII-D plasma, shot 169286, during flattop. In this shot, the plasma was translated vertically up and down several times, providing the basis for a vertical-displacement scan. We used the time slice at 3183 ms as the reference equilibrium because the magnetic axis was close to the machine midplane, with $Z_{axis} = 0$. This is the only equilibrium used in the present paper. Choosing a midplane-centred time slice is also advantageous because the DIII-D Thomson scattering diagnostic measures the electron density at locations close to the midplane, giving a better constrained density profile for the equilibrium used in \textit{Scotty}. The magnetic equilibrium and electron density profile from this time slice were therefore processed with OMFIT \cite{Meneghini:OMFIT:2015, Pratt:OMFIT:2025} and used as inputs to the synthetic diagnostic.

\begin{figure}
    \includegraphics[width=0.45\textwidth]{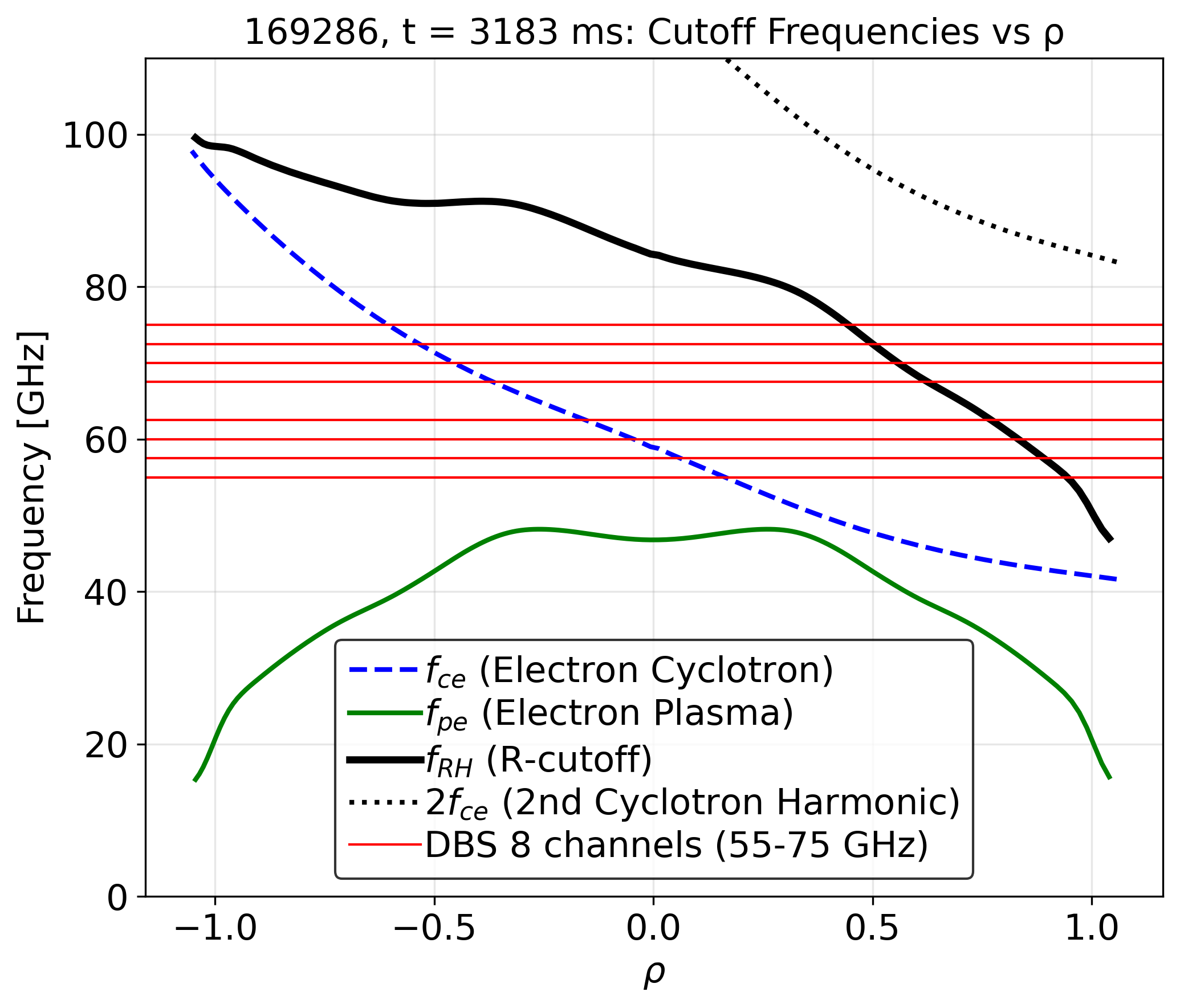}
    \caption{Characteristic frequencies for the reference DIII-D equilibrium used in this study. The V-band DBS frequencies considered here access X-mode cutoff locations from near the last-closed flux surface to the plasma core.}
    \label{fig:cutoff_freqs}
\end{figure}

While DIII-D has several DBS diagnostics, we used the properties of the V-band ($55$--$75$ GHz) system described in previous work\cite{Rhodes:DBS_quasioptical:2010} as input for our simulations. This system had one-dimensional steering; it could steer poloidally but not toroidally. In our simulations, we took the diagnostic to operate in X-mode, consistent with the actual experimental configuration. For this plasma, the corresponding cutoff locations span from near the last-closed flux surface to the core, as shown in Fig.~\ref{fig:cutoff_freqs}.

We seek to compare two methods of varying the measured turbulence wavenumber. 
\begin{itemize}
    \item \textbf{Vertical-displacement scan}. The plasma is held fixed while the DBS launch position is displaced vertically. This is equivalent to the fixed DBS launch but vertically displacing plasma in experiments, but less cumbersome to simulate. The vertical-displacement scan range is chosen to be the same as that achieved in shot 169286, from approximately $Z_{\rm ant}=0$ m  to $Z_{\rm ant}=0.20$ m. The poloidal angle was fixed at $-3^\circ$. Note that the negative sign corresponds to an upward launch.
    \item \textbf{Poloidal-angle scan}. The plasma equilibrium is held fixed while the poloidal launch angle is varied. The poloidal launch angle was varied between $-2^\circ \geq \varphi_p \geq -14^\circ$, chosen so that the measured turbulence wavenumbers approximately match those accessed by the vertical-displacement scan.  The DIII-D DBS system is experimentally unable to access this entire range of poloidal launch angles due to limitations of the port window size, but this does not constrain our current work, which is to illustrate the effect of such a scan. The launch position was fixed at $Z_{\rm ant}=-0.0157$ m, which is the actual launch position of the reference DBS system. 
\end{itemize} 
In both scans, the launch frequencies, toroidal launch angle, polarisation, and input plasma equilibrium are otherwise held fixed to enable an apples-to-apples comparison of the two methods.

\begin{figure}
    \includegraphics[width=0.45\textwidth]{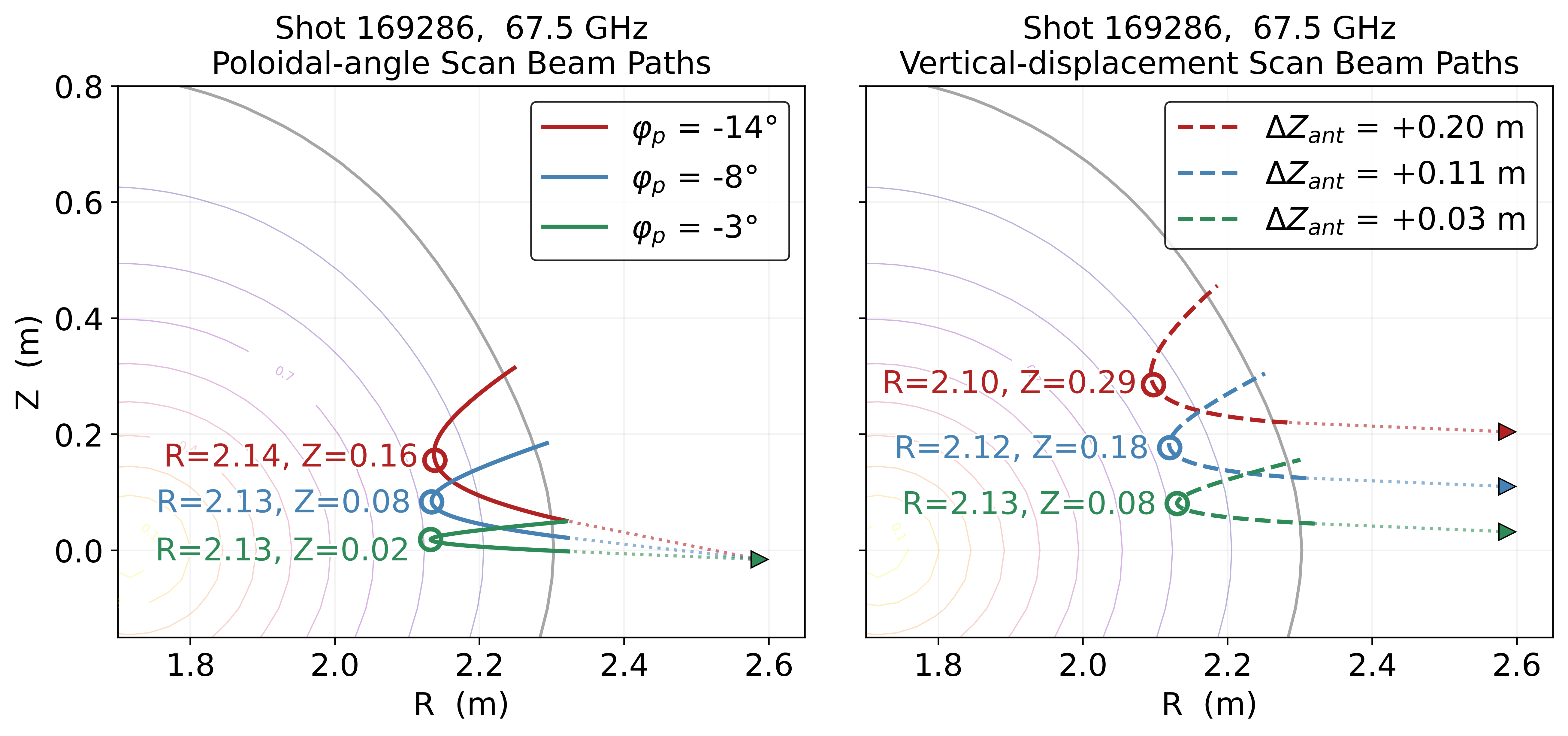}
    \caption{Representative DBS beam trajectories for the vertical-displacement and poloidal-angle scans. Note the difference in vertical height and radial location of the cutoff location when comparing the two methods at the equivalent launch positions.}
    \label{fig:beam_paths}
\end{figure}

For each frequency and each point in the vertical-displacement and poloidal-angle scans, we use \textit{Scotty} to trace the DBS probe beam through the plasma, see Figure \ref{fig:beam_paths}. 

The inputs to \textit{Scotty} are the magnetic equilibrium, electron density profile, launch position, beam width and curvature at launch, launch frequency, polarisation, and launch angles. \textit{Scotty} then calculates the central-ray trajectory, the probe-beam wavevector along that trajectory, and the beam properties needed to evaluate the DBS filter function. The filter function has three main parts: the probe beam's electric field, mismatch attenuation, and polarisation effects, of which the former two are the most important \cite{Hall-Chen:DBS_beammodel:2022}. Moving forward, we define the nominal cutoff location as the point along the central ray where the probe beam's wavenumber is minimised. While the backscattered DBS signal is usually approximated to predominantly come from scattering near the cutoff, we do not assume this \emph{a priori}. The primary fluctuation wavenumber responsible for backscattering is determined via the Bragg condition,
\begin{equation}
    \mathbf{k}_\perp \simeq -2 \mathbf{K},
\end{equation}
where $\mathbf{K}$ is wavevector of the probe-beam. This is evaluated at every point along the central ray, accounting for scattering at points away from the cutoff location. 

We then use the electric field calculated from beam tracing, together with the reciprocity theorem, to determine the synthetic backscattered power. This is also done by \textit{Scotty}. In this study, we assume that the statistical properties of the turbulent density fluctuations do not vary with radial or poloidal position. Hence, the backscattered spectral density is given by
\begin{equation} \label{eq:spectral_power}
    p_r \left( \omega \right) \propto 
    \int F_i \left( l \right)
    \delta \tilde{n}_e^2 
    \left[ \mathbf{k}_\perp (l), \omega \right]
    \textrm{d}l .
\end{equation}
Here, $p_r$ is the backscattered spectral density, $l$ is the arc length along the central ray, $F_i$ is the instrumentation or filter function, and $\delta \tilde{n}_e^2$ is the density-fluctuation spectrum evaluated at the wavenumber selected by the Bragg condition.  $\omega$ contains both the turbulence phase velocity and background plasma flow velocity. In this paper, we are concerned with the total backscattered power rather than its spectral density. Hence, $\omega$ is integrated over, and the exact flow profile does not affect our results. The filter function contains the diagnostic response, including the effects of the probe-beam electric field, beam width, wavefront curvature, and matching between the probe beam and the local magnetic field \cite{Hall-Chen:DBS_beammodel:2022}. For instance, regions where the beam electric field is larger contribute more strongly to the received signal, even if the underlying turbulent fluctuation amplitude is the same.\cite{Ruiz:beam:2025} 

To illustrate the effect of the fluctuation spectrum, we assume an ion-temperature-gradient-driven density-fluctuation spectrum using the analytic form fitted to previous gyrokinetic simulations\cite{Ruiz:RCDR:2022},
\begin{equation} \label{eq:delta_ne}
    \delta \tilde{n}_e^2 \propto \left[ 1 
    + \left( \frac{ \left| k_n \rho_s \right| }{0.17} \right)^{3.9}
    + \left( \frac{ \left| k_b \rho_s - 0.24 \right| }{0.28} \right)^{3.24}
    \right]^{-1} .
\end{equation}
Here, $k_n$ is the component of the turbulence wavevector, $\mathbf{k}_\perp$, normal to the flux surface, $k_b$ is that in the flux surface and perpendicular to the equilibrium magnetic field, and $\rho_s$ is the ion sound gyroradius. This provides a believable anisotropic spectrum with which to compare how vertical-displacement and poloidal-angle scans sample wavenumber space. The synthetic spectral power, and thus the backscattered power, is then obtained by evaluating equation (\ref{eq:spectral_power}) with the fluctuation spectrum in equation (\ref{eq:delta_ne}) at all positions within the plasma. Note that in experimental investigation of the wavenumber spectrum, we would do the reverse --- measure the backscattered power, invert the problem, and find the form of the density fluctuation spectrum. However, this is beyond the scope of present work.

\section{Backscattered power} \label{sec:results} 
We now compare vertical-displacement and poloidal-angle scans using the \textit{Scotty} synthetic diagnostic. The comparison is carried out in two steps. First, we compare the measurement locations and fluctuation wavenumbers selected by the two scans, using the nominal cutoff as a compact description of where and what each DBS channel measures. Secondly, we compare the predicted backscattered power, which includes additional diagnostic effects such as the beam's electric field and mismatch attenuation. This distinction is important: two scans may access similar cutoff locations and wavenumbers, but still produce different received powers because the diagnostic response is not the same.

\subsection{Dominant measurements location and wavenumbers}
We first compare the dominant measurement locations and wavenumbers accessed by the vertical-displacement and poloidal-angle scans. The scan ranges were chosen so that the two methods access similar values of the probe-beam wavenumber at cutoff, $K_c$, normalised to the vacuum wavenumber, $K_0$. This normalisation allows the comparison to be made across the full V-band frequency range. In both cases, we restrict the comparison to $0.05 < K_c/K_0 < 0.4$, corresponding to the usual DBS regime \cite{Ruiz:beam:2025}, and to cutoff locations above the midplane.

The dominant measurement location is taken to be the nominal cutoff, defined as the point along the central ray where the probe-beam wavenumber is minimised. This does not require the entire backscattered signal to originate exactly at the cutoff. Rather, the cutoff provides a useful first comparison because DBS signals are usually dominated by scattering from near this location. The full synthetic backscattered power, including the beam and instrumentation effects along the ray, is considered in the next subsection.

\begin{figure}
    \includegraphics[width=0.45\textwidth]{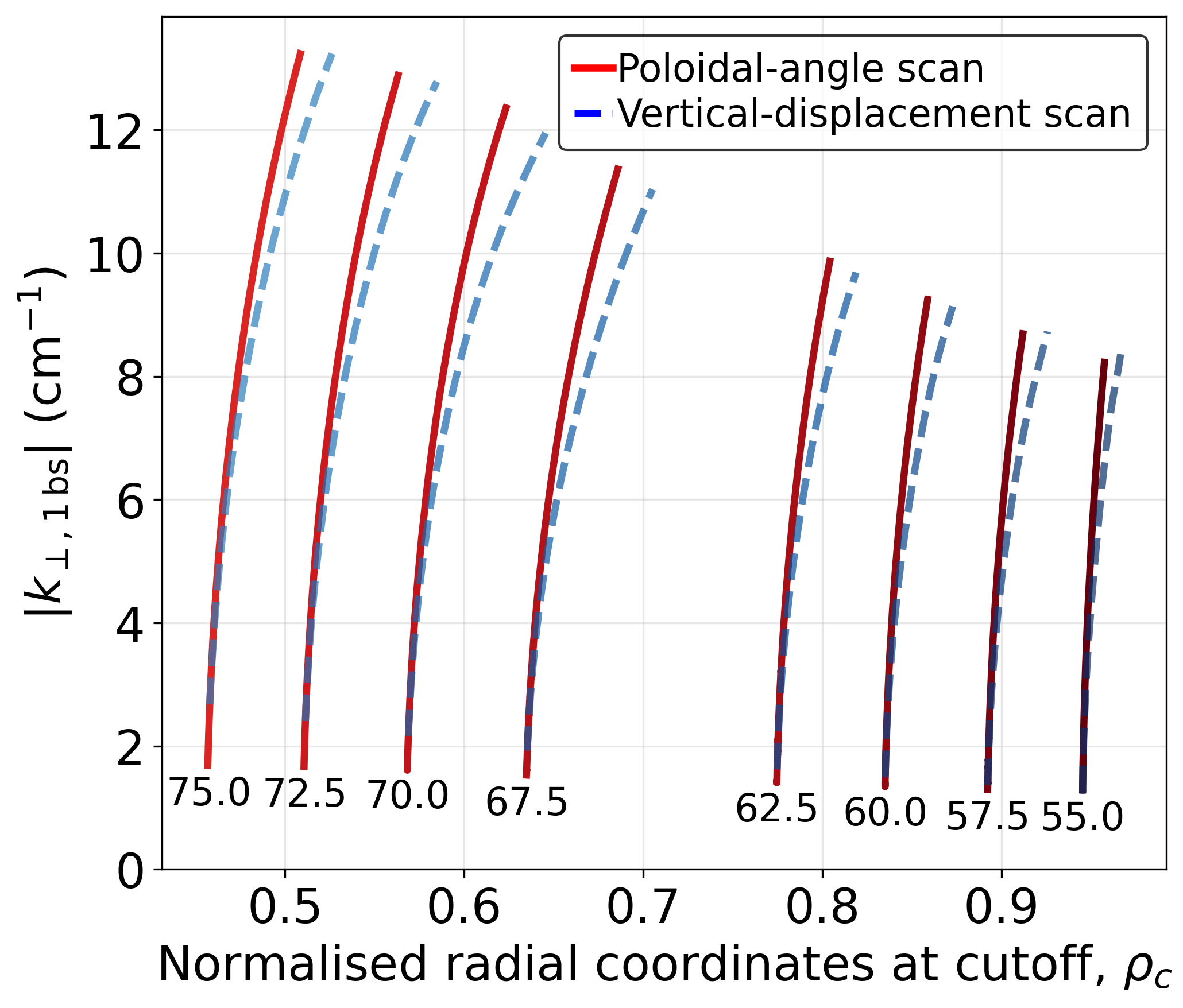}
    \caption{Measured fluctuation wavenumbers and cutoff locations for the poloidal-angle and vertical-displacement scans. The scan ranges were chosen to access similar values of $K_c/K_0$, and hence similar measured wavenumbers. The two scans agree well at low wavenumber. At higher wavenumber, small systematic differences appear. For a given frequency, the vertical-displacement scan measures a slightly larger wavenumber at low $K_c$, while at high $K_c$ it reaches similar wavenumber at slightly larger $\rho_c$.}
    \label{fig:kperp_rhoc}
\end{figure}

We found that the two scans access broadly similar measured turbulence wavenumbers over the full frequency range, see Figure~\ref{fig:kperp_rhoc}. 
That is, a vertical-displacement scan can approximately reproduce the radial and wavenumber coverage of a conventional poloidal-angle scan. However, the agreement is not exact. As frequency increases, two systematic differences appear. First, at lower values of $K_c$, the vertical-displacement scan gives a slightly larger measured wavenumber than the poloidal-angle scan at the same frequency. Secondly, at larger values of $K_c$, the two scans access similar measured wavenumbers, but the vertical-displacement scan does so at a slightly larger normalised radius, further towards the plasma edge.

\begin{figure}
    \includegraphics[width=0.45\textwidth]{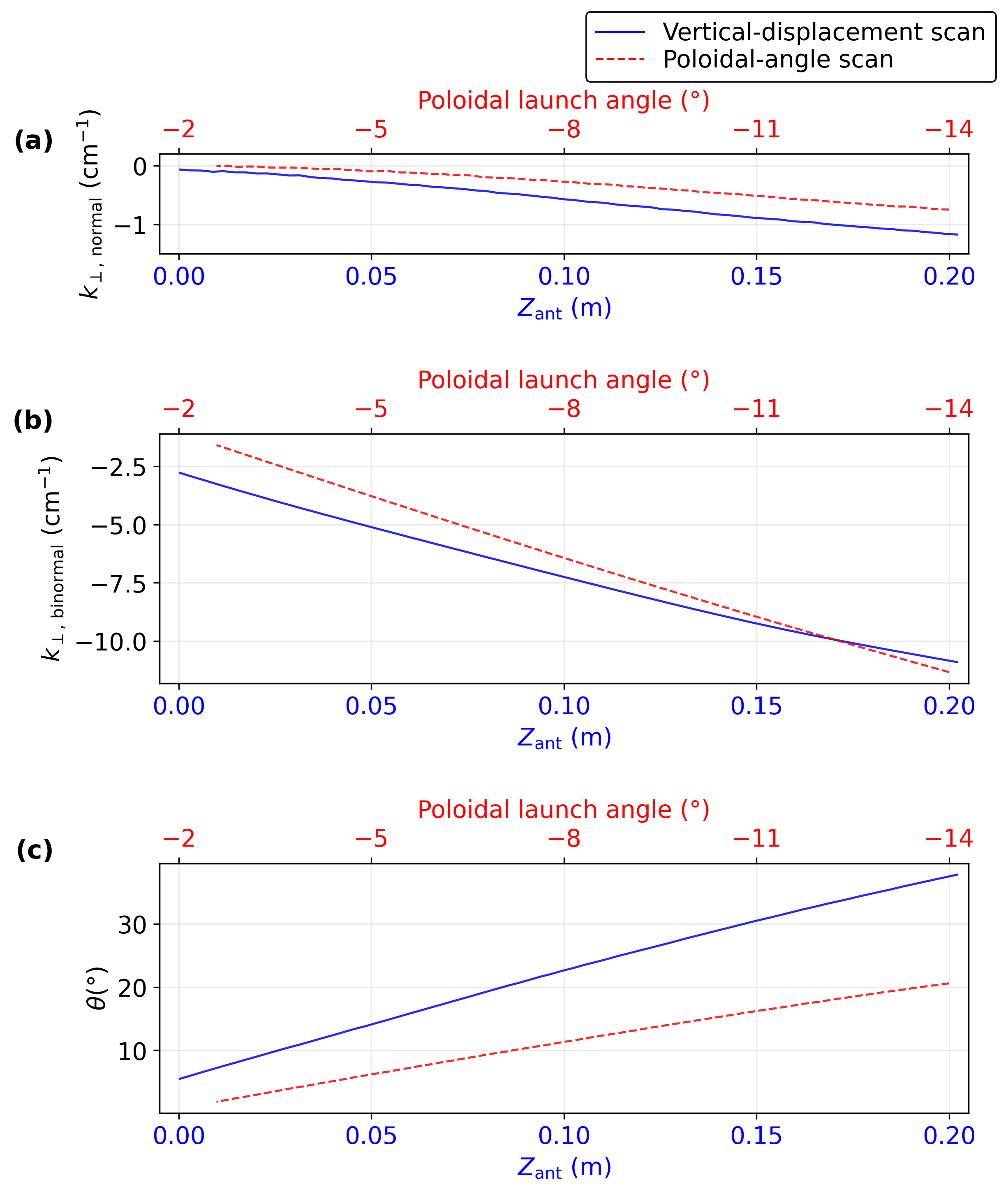}
    \caption{Cutoff locations and measured wavenumber components for the poloidal-angle and vertical-displacement scans. (a) Normal component of the measured fluctuation wavevector, $k_n$, and (b) binormal component, $k_b$, which, while broadly similar between the two scans, remain systematically different. (c) Poloidal location of the cutoff differs more strongly, indicating that the two methods can access similar radial locations and total wavenumbers while sampling different poloidal regions of the plasma.}
    \label{fig:resolved_quantities}
\end{figure}

To understand these differences in more detail, we compared the full 2D cutoff position and both components of the measured fluctuation wavevector. First, in a real device, turbulence is expected to vary spatially with both radial and poloidal locations. Secondly, as turbulence is anisotropic, the total measured wavenumber might not be sufficient to determine whether the two scans sample the same region of wavenumber space. Hence, we decomposed the measured wavevector into normal and binormal components, see equation (\ref{eq:delta_ne}).
We found that the normal and binormal components of the turbulence wavevector were indeed similar, though not exactly the same, between the two scans, see Figure~\ref{fig:resolved_quantities}. 
The largest difference in $k_n$ occurs at the largest wavenumber, where the vertical-displacement scan gives $k_n=-1.18 \textrm{cm}^{-1}$ compared to $-0.75\textrm{cm}^{-1}$ for the poloidal-angle scan. Conversely, the largest difference in $k_b$ occurs at the lowest wavenumber, where the vertical-displacement scan gives $k_b=-2.78\textrm{cm}^{-1}$ compared to $-1.53\textrm{cm}^{-1}$ for the poloidal-angle scan. These differences are modest compared with the overall agreement between the two scans, but they show that the two methods do not sample precisely the same region of wavenumber space.

The most significant difference between the two scans is in the poloidal position of the cutoff. Although the two scans access similar radial locations, the vertical-displacement scan reaches cutoff at poloidal locations that can be roughly twice as far from the midplane as those of the poloidal-angle scan. If the turbulence amplitude is approximately poloidally invariant, the two scan methods should be comparable. However, in real plasmas, especially where turbulence is ballooning-like or otherwise poloidally localised, a vertical-displacement scan may sample a different fluctuation amplitude even when the radial location and measured wavenumber are similar.

\subsection{Backscattered power spectrum}
Similar cutoff locations and measured wavenumbers do not necessarily imply similar received power. The DBS signal also depends on the diagnostic filter function, including the probe-beam's electric field and mismatch attenuation. We therefore compare the synthetic backscattered power for three cases 
\begin{itemize}
    \item A poloidal-angle scan with no toroidal steering, as was the case with the poloidal scans done thus far in the paper,
    \item A poloidal-angle scan in which the toroidal launch angle is optimised to minimise mismatch at cutoff
    \item The vertical-displacement scan with fixed launch geometry 
\end{itemize}
They all have the same turbulence spectrum given in equation (\ref{eq:delta_ne}). As we shall later see, mismatch attenuation is the dominant instrumentation effect at high k. As such, this comparison enables us to separately consider the effect of the other instrumentation functions on the backscattered power.

\begin{figure}
    \includegraphics[width=0.45\textwidth]{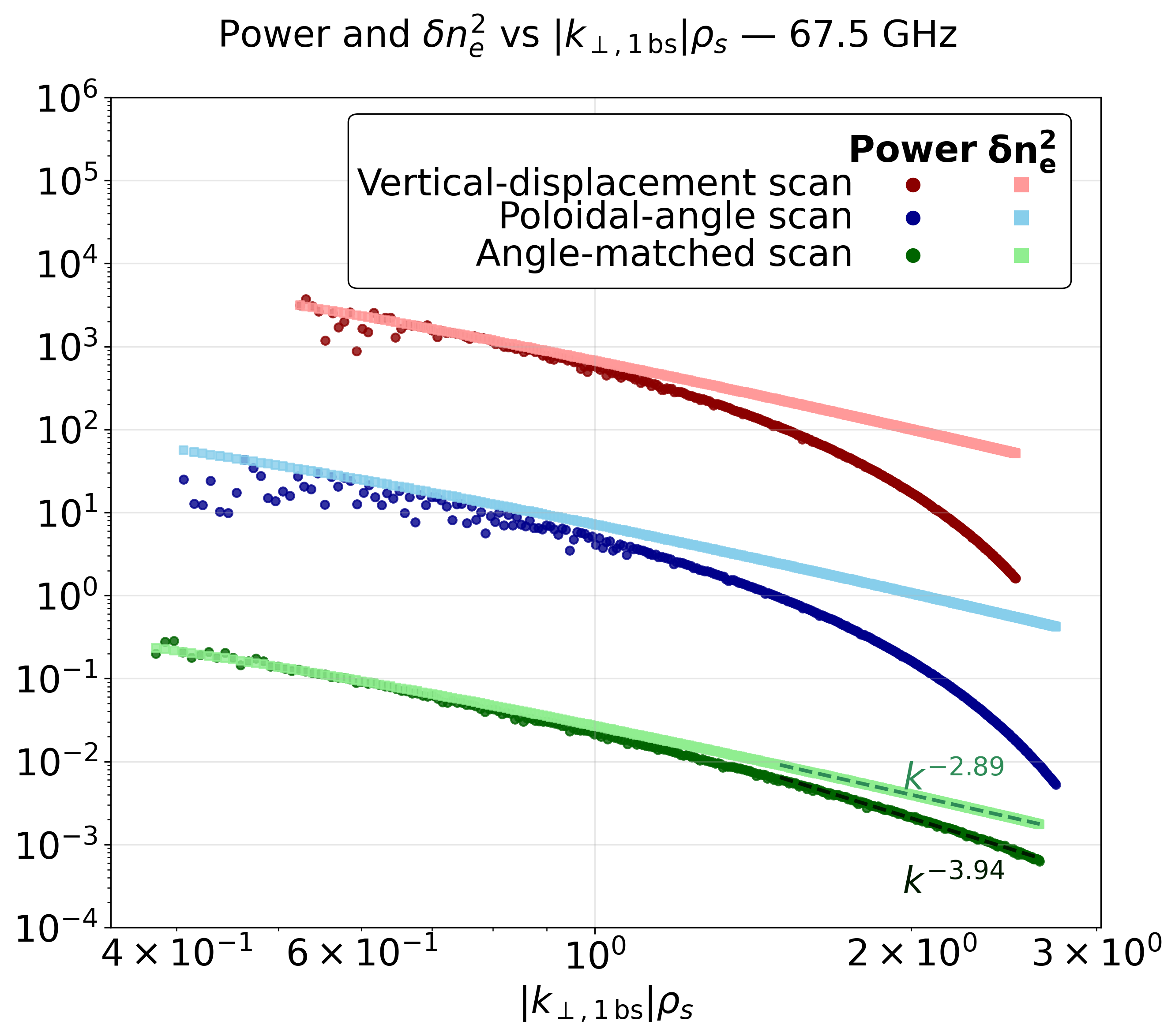}
    \caption{Comparison of synthetic backscattered power for the poloidal-angle and vertical-displacement scans, as predicted by \textit{Scotty}. The assumed turbulent density spectrum at the cutoff location is also shown, for the given measured $k$ at cutoff from the Bragg condition. All curves are normalised to compare their spectral shape. Both the vertical-displacement scan and poloidal-angle scan show a strong fall-off in received power at high wavenumber than the assumed density spectrum. This additional fall-off is mainly caused by mismatch attenuation rather than by the turbulence spectrum alone.}
    \label{fig:spectrum}
\end{figure}

Using equation (\ref{eq:spectral_power}), we calculated the predicted backscattered power for each of the three cases, see Figure \ref{fig:spectrum}. 

Both the vertical-displacement scan and poloidal-angle scans show a strong fall-off in received power at high wavenumber. This fall-off is much steeper than that of the imposed density-fluctuation spectrum, indicating that it is not simply a consequence of the assumed $k$-spectrum. Instead, the comparison with the toroidally optimised poloidal-angle scan shows that the additional reduction is primarily caused by mismatch attenuation. A vertical-displacement scan without toroidal steering therefore requires synthetic-diagnostic modelling of mismatch attenuation before the measured backscattered power can be interpreted as a turbulence spectrum.

Even after mismatch is minimised, the backscattered power does not have exactly the same large-wavenumber scaling as $\delta \tilde{n}_e^2$. At large k, the difference between the gradients, for all frequencies, is around $k^{-1.0}$. This is expected because the received power still contains non-mismatch instrumentation effects, including the variation of beam width and wavefront curvature. A similar difference between the underlying density-fluctuation spectrum and the measured DBS power spectrum was found by Pratt \textit{et al.}~\cite{Pratt:DBS:2024}. Therefore, even in the matched case, the measured power spectrum should not be directly interpreted as the density-fluctuation spectrum without a synthetic diagnostic.

\section{Conclusions}
We evaluated vertical-displacement scan as a route to single-shot DBS measurements of turbulence wavenumber spectra without special DBS hardware for in-shot steering. In the conventional approach, the measured wavenumber is varied by changing the poloidal launch angle, usually between repeated discharges. This places a strong requirement on shot-to-shot repeatability. In the vertical-displacement scan approach considered here, the DBS launch geometry is held fixed while the relative vertical position of the antenna and plasma is changed. As such, wavenumber scans that would normally require repeated shots can instead be obtained within a single discharge.

Using a DIII-D L-mode equilibrium and the \textit{Scotty} synthetic diagnostic, we found that vertical-displacement and poloidal-angle scans can access broadly similar radial locations and measured fluctuation wavenumbers. This establishes vertical-displacement scan as a viable approach for DBS measurements of the turbulent density fluctuation k-spectrum. The two scans have certain differences in what and where they measure. The normal and binormal components of the measured wavenumber differ modestly, and the largest difference is in the poloidal position of the cutoff location. In the case studied here, the vertical-displacement scan sampled cutoff locations around twice farther from the midplane than the poloidal-angle scan. This would be important in plasmas where the turbulence amplitude has strong poloidal variation, but it is a calculable difference rather than a fundamental obstacle.

We also compared the synthetic backscattered power from the two methods using the same assumed density-fluctuation spectrum. Both scans showed a strong fall-off in backscattered power at high-wavenumber, mainly due to mismatch attenuation rather than the imposed turbulence spectrum. We note that the mismatch attenuation is larger for the vertical-displacement scan. Even after toroidal matching was optimised, the backscattered power did not exactly follow the assumed density-fluctuation spectrum because non-mismatch instrumentation effects, such as beam width and wavefront curvature still contributed. This emphasises the need for a synthetic diagnostic when interpreting DBS spectra.

In devices where the plasma can be vertically translated, the vertical-displacement scan method reduces reliance on repeated discharges and enables a single-shot wavenumber scan. Conversely, in devices where the plasma must remain fixed, the same principle can be implemented by translating the DBS launch assembly vertically. This may be attractive on devices such as J-TEXT, where vertical motion of the assembly can be more practical than implementing poloidal steering because of space constraints. The downside is that the measurement geometry can change in a more complicated way, especially through mismatch attenuation and poloidal variation. These effects can be calculated with \textit{Scotty} and included in the analysis. With this modelling in place, vertical-displacement scan offers a practical route to validating models of turbulence  over a range of radial locations and turbulent wavenumbers.

\begin{acknowledgments}
This work was partially funded by A*STAR, via the Strategic Tokamak Research for Industrial Deployment, and Energy (STRIDE) grant [H26-MSE1527], a Young Individual Research Grant [M23M7c0127], and a SERC Central Research Fund. This material is also based upon work supported by the U.S. Department of Energy, Office of Science, Office of Fusion Energy Sciences, using the DIII-D National Fusion Facility, a DOE Office of Science user facility, under Awards DEFC02-04ER54698 and DE-SC0019352. 

\textbf{Disclaimer}.
This report was prepared as an account of work sponsored by an agency of the United States Government. Neither the United States Government nor any agency thereof, nor any of their employees, makes any warranty, express or implied, or assumes any legal liability or responsibility for the accuracy, completeness, or usefulness of any information, apparatus, product, or process disclosed, or represents that its use would not infringe privately owned rights. Reference herein to any specific commercial product, process, or service by trade name, trademark, manufacturer, or otherwise does not necessarily constitute or imply its endorsement, recommendation, or favoring by the United States Government or any agency thereof. The views and opinions of authors expressed herein do not necessarily state or reflect those of the United States Government or any agency thereof. \\
\textbf{AI declaration}. 
ChatGPT was used to assist with language editing, manuscript structure, and clarity of presentation. This included drafting and refining wording for selected sections, figure captions, and the conclusion. The scientific concept, simulations, data analysis, figures, interpretation, and final technical content were produced without AI. All AI-assisted text was reviewed and edited by the authors before inclusion.\\
\textbf{Data availability.}
The data that support the findings of this study are available from the corresponding author upon reasonable request.
\end{acknowledgments}

\nocite{*}
\bibliography{aipsamp}

\end{document}